\documentclass{aa}  
\usepackage[switch, modulo]{lineno}
\usepackage{amsmath}
\usepackage{tablefootnote}
\usepackage{bm}
\usepackage{xcolor}
\usepackage[normalem]{ulem}
\usepackage{graphicx}
\usepackage{txfonts}
\usepackage[colorlinks,linkcolor=blue,citecolor=blue,linktocpage=true,breaklinks, plainpages=false,urlcolor=blue]{hyperref}

\usepackage{rotating}
\usepackage{sidecap}
\usepackage{subcaption}
\usepackage{url}
\usepackage{soul}
\hypersetup{allcolors=blue}
\usepackage{float}
\usepackage[tracking=true,kerning=true,shrink=50]{microtype}
\usepackage{lipsum}
\usepackage{txfonts}
\usepackage[export]{adjustbox} % For centering the images
\bibpunct{(}{)}{;}{a}{}{,}
\DeclareUnicodeCharacter{00A0}{~}
\usepackage[us,12hr]{datetime}
\usepackage{ulem}

\renewcommand{\jobname}{spectralres}

\begin{document} 

    \title{Spectral resolution effects on the information content in solar spectra}
    % \authorrunning{}
    
    \author{C. J. D\'{i}az Baso
        \inst{1,2}
        \and
        I. Mili\'{c}
        \inst{3,4}
        \and
        L. Rouppe van der Voort
        \inst{1,2}
        \and
        R. Schlichenmaier
        \inst{3}
        }
    
    \institute{
    Institute of Theoretical Astrophysics,
    University of Oslo, %
    P.O. Box 1029 Blindern, N-0315 Oslo, Norway
    \and
    Rosseland Centre for Solar Physics,
    University of Oslo, % 
    P.O. Box 1029 Blindern, N-0315 Oslo, Norway
    \and
    Institute for Solar Physics (KIS), Georges-Köhler-Allee 401a
79110 Freiburg, Germany
    \and
    Faculty of Mathematics, University of Belgrade, Studentski Trg 16, Belgrade, Serbia
    \\
    \email{carlos.diaz@astro.uio.no}
    }

   % \date{Received ; accepted }
   \date{Draft: compiled on \today\ at \currenttime~UT}
   \authorrunning{D\'iaz Baso et al.}

% \abstract{}{}{}{}{} 
% 5 {} token are mandatory
 
    \abstract
    {When interpreting spectropolarimetric observations of the solar atmosphere, wavelength variations of the emergent intensity and polarization translate into information on the depth stratification of physical parameters such as temperature, velocity, and magnetic field. Resolving the fine details in the shapes of the spectral lines and their polarization gives us the capability to resolve small-scale depth variations in these physical parameters. With the advent of large-aperture solar telescopes and the development of state-of-the-art instrumentation, the requirements on spectral resolution have become a prominent question.} 
    {We aim to quantify how the information content contained in a representative set of polarized spectra of photospheric spectral lines depends on the spectral resolution and spectral sampling of that spectrum.}
    {We use a state-of-the-art numerical simulation of a sunspot and the neighboring quiet Sun photosphere to synthesize polarized spectra of magnetically sensitive neutral iron lines. We then apply various degrees of spectral degradation to the synthetic spectra and analyze the impact on its dimensionality using Principal Component Analysis (PCA) and wavelength power spectrum using wavelet decomposition. Finally, we apply the Stokes Inversion based on Response functions (SIR) code to the degraded synthetic data, to assess the effect of spectral resolution on the inferred parameters.}
    {
    We find that the dimensionality of the Stokes spectra and the power contained in the small spectral scales significantly change with the spectral resolution. We find that regions with strong magnetic fields where convection is suppressed have more homogeneous atmospheres and produce less complex Stokes profiles. On the other hand, regions with strong gradients in the physical quantities give rise to more complex Stokes profiles that are more affected by spectral degradation. The degradation also makes the inversion problem more ill-defined, so inversion models with a larger number of free parameters overfit and give wrong estimates.
    }
    {
    The impact of spectral degradation in the interpretation of solar spectropolarimetric observations depends on multiple factors, including spectral resolution, noise level, line spread function (LSF) shape, complexity of the solar atmosphere, and the degrees of freedom in our inversion methods. To mitigate this impact, incorporating a good estimation of the LSF into the inversion process is recommended. Having a finely sampled spectrum may be more beneficial than achieving a higher signal-to-noise ratio per wavelength bin. Considering the inclusion of different spectral lines that can counter these effects, and calibrating the effective degrees of freedom in modeling strategies, are also important considerations. These strategies are crucial for the accurate interpretation of such observations and have the potential to offer more cost-effective solutions.
    }
    
    \keywords{Sun: photosphere -- Line: formation  -- Methods: observational -- Sun: magnetic fields -- Radiative transfer}
    
    \maketitle
   
%
% -----------------------------------------------------------------------
\section{Introduction}

Physical processes taking place in the solar atmosphere exhibit a remarkable diversity of spatial, temporal, and energetic scales, necessitating measurements with high spatial, spectral, and temporal resolution, and sufficient signal-to-noise ratio (SNR) to comprehend their underlying nature \citep[see, e.g.,][]{Iglesias2019}. Specifically, resolving the wavelength variations of the emergent intensity and its polarization gives us insight into the depth variation of physical parameters such as the temperature, the velocity, and the magnetic field \citep{1996SoPh..164..169D}. This is because the absorption and emission processes in spectral lines show strong variations in wavelength, giving us access to a range of depths in the atmosphere of the Sun while observing a relatively narrow wavelength range. 

Two classes of instruments are mainly used to spectrally resolve the light received at the telescope: filtergraph- and spectrograph-based systems. Filtergraphs, with the Fabry-P\'erot interferometer being the most popular choice, are used for narrow-band imaging. They obtain pseudo-monochromatic images of the observed field of view (FoV). These instruments brought a revolution to high-resolution solar spectropolarimetry \citep[e.g.][]{CRISP,2006SoPh..236..415C}, making it possible to observe and analyze very small (<100\,km) spatial details in the solar atmosphere \citep[e.g.][]{Rouppe2017ApJ, DiazBaso2021A&A...647A.188D} and study their spatial distribution in large FoVs \citep[e.g.][]{Kianfar2020A&A...637A...1K, Morosin2022A&A...664A...8M}. Spectral fidelity (the degree of similarity between the original spectrum and the recorded one) of these instruments is limited and often comes at the cost of sacrificing the signal-to-noise ratio and/or temporal resolution \citep[e.g.][]{RolfFP2023, DiazBasoFP2023}, making resolving complicated spectral lines in detail borderline unfeasible. On the other hand, slit-based spectrographs (from now on, spectrographs) trade one spatial dimension for the instantaneous wavelength information, traditionally through a spectrograph slit \citep[e.g.][]{gris2012}. This way, the entire spectrum is captured with high spectral detail, but to obtain a two-dimensional map of an extended source, the slit needs to be moved to scan the FoV. Spectrographs are also used as wavelength discriminators in integral field units like MiHi \citep{MiHiI2022, Rouppe2023A&A}, which are capable of capturing the spatial and wavelength information simultaneously, albeit at a limited field of view. 

The high spectral resolution and fine sampling result in an increased number of measurements and, consequently, an improved SNR. Furthermore, resolving fine spectral features allows us to probe complicated depth variations of physical parameters \citep{1992ApJ...398..359S}. This has been important for studies like probing the magnetic fields in the quiet Sun \citep{Marian2009ApJ...700.1391M}, filaments and prominences \citep{DiazBaso2016ApJ...822...50D, DiazBaso2019A&A...625A.128D}, sunspots \citep{2007ApJ...666L.133B, 2024arXiv240807645E} and magnetic flux-emerging regions \citep{Yadav2019A&A...632A.112Y}. Thus spectrographs are common choices for instrument suites of ground- and space-based telescopes like DST/SPINOR \citep{SocasNavarro2006SoPh}, ZIMPOL \citep{Povel2001ASPC..248..543P}, the GREGOR Infrared Spectrograph \citep[GRIS;][]{gris2012} at the GREGOR telescope \citep{gregor2012}, the TRI-Port Polarimetric Echelle-Littrow spectrograph \citep[TRIPPEL;][]{Kiselman2011A&A} at the Swedish 1-m Solar Telescope \citep[SST;][]{Scharmer2003}, the spectropolarimeter (SP) of the Hinode satellite \citep{Kosugi2007SoPh}, among others. They remain crucial instruments for providing high-fidelity spectral information at the next-generation 4-meter class of telescopes, such as the existing \textit{Daniel K. Inouye} Solar Telescope \citep[DKIST;][]{DKIST2020} with the Visible Spectro-Polarimeter \cite[ViSP;][]{deWijn2022} and the planned European Solar Telescope \citep[EST;][]{QuinteroNoda2022} with Integral Field Units like MiHi \citep{MiHiI2022} or MuSICa \citep{2022JAI....1150014D}.

Choosing the spectral resolution of a spectrograph system is a challenge: an increase in the spectral resolution necessitates using finer sampling and thus decreasing the SNR per spectral bin, as well as the wavelength range of the observations. High spectral resolution increases the design complexity, which increases the cost of the equipment and compromises the robustness of the instrument. On the other hand, the loss of the spectral resolution poses a potential loss of information which can ultimately cause us to miss important physical content or to misdiagnose physical conditions.

This work investigates the impact of limited spectral resolution on the information content in spectropolarimetric measurements and inferred quantities. Our goal is to quantify how the information content contained in a representative set of polarized spectra of photospheric spectral lines depends on the spectral resolution and spectral sampling of the instrument used to acquire that spectrum. We achieve this by calculating synthetic spectra from a state-of-the-art MHD simulation of the solar atmosphere and degrading them according to different spectral resolutions. We then analyze the information content in the original and degraded spectra and in the atmospheric stratification inferred from them using a spectropolarimetric inversion code. This strategy was earlier successfully executed by for example \citet{delaCruz2012A&A...543A..34D, Millic2019A&A...630A.133M, Campbell2021A&A...654A..11C, Inv_sim_CQ}. We believe these works are excellent references to understand the impact of the observation mode and analysis tools, optimizing observational strategies, identifying instrumental requirements, and refining our scientific interpretation. This work is organized as follows: we begin with a brief introduction and presentation of the synthetic observables we use in this work (Sect.~\ref{sec:data}). We then analyze the dimensionality of the data to quantify the complexity of the Stokes profiles (Sect.~\ref{sec:pca}), followed by an analysis of the spectral scales present in the observables (Sect.~\ref{sec:wavelet}) and finalizing with a study of the accuracy of the inferred atmospheric parameters from spectropolarimetric inversions under different levels of instrumental degradation (Sect.~\ref{sec:inversions}). Finally, we discuss these discrepancies and present our conclusions and recommendations for the instrument design (Sect.~\ref{sec:conclusions}).

% -------------------------------------------------------
\section{Data preparation}\label{sec:data}
% -------------------------------------------------------

% -------------------------------------------------------
\subsection{Synthetic data}
% -------------------------------------------------------
The main reason for opting for high spectral resolution observations is to improve the ability to resolve the shapes of spectral lines, allowing for a detailed inference of the vertical stratification of physical quantities. For a given spectral resolution, narrower spectral lines will be the most affected. In solar conditions, narrow spectral lines are typically the lines of heavier elements formed in the solar photosphere. Therefore, we chose to study the two magnetically sensitive spectral lines of neutral iron around 630\,nm, which are also observed by the space-based slit spectropolarimeter \citep{2013SoPh..283..579L} onboard the Hinode \citep{Hinode} spacecraft, and a common choice for ground-based observations, e.g. ViSP at DKIST. To calculate the synthetic spectra we used a radiative-magneto-hydrodynamic (RMHD) simulation of a sunspot performed with the MURaM code \citep{Vogler2005A&A,Rempel2017ApJ,2021A&A...656A..92S}. The sunspot also contains a quiet (weakly magnetized) solar atmosphere around, thus providing a diverse set of observables that can be found in real observation. That is, we find strong and weak magnetic fields of various inclinations as well as regions with varying velocities and temperatures. The size of the simulation box considered for the analysis is $ 2048\times 256\times 256$ in $(x,y,z)$ with a grid size of $\Delta x = 20$, $\Delta y = 20$, and $\Delta z = 8$ km. This simulation is considered to be the state of the art in the generation of a numerical solar sunspot and surroundings \citep{Tiwari2013A&A} and it has been used to test different spectropolarimetric inversion approaches \citep{AsensioRamos2019A&A, PastorYabar2019A&A}. 
% The original size of the simulation box is 4096×512×256 with a grid size of ∆x = 10, ∆y = 10, and ∆z = 8 km.

\begin{figure*}[htp!]
\centering
\includegraphics[width=1.0\textwidth]{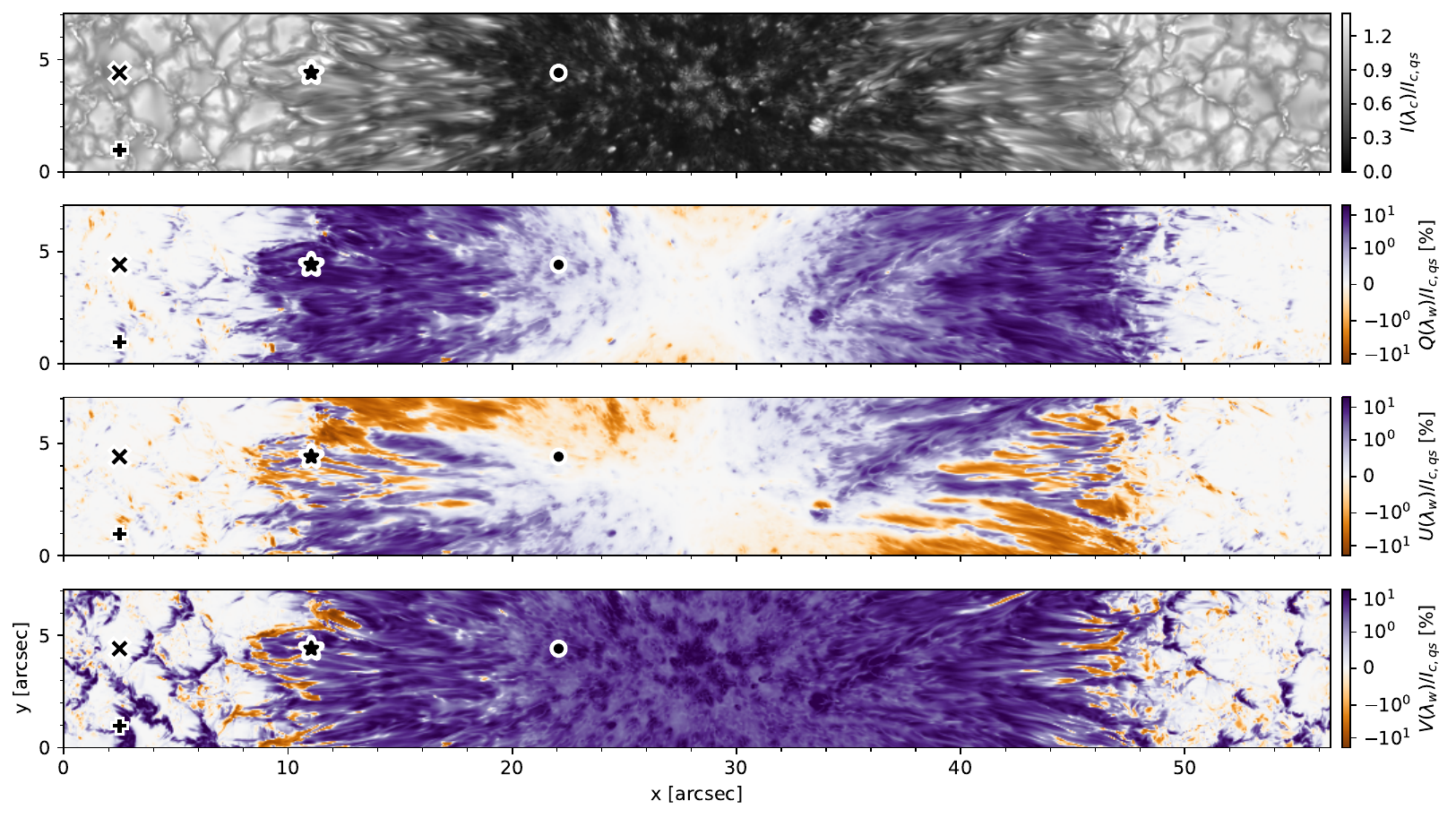}
\caption{Maps of synthetic intensity and polarization calculated from a snapshot of the MURaM simulation of a sunspot. The upper panel shows the continuum intensity, and the rest display the Stokes $Q$, $U$, and $V$ signals at $\lambda_w=6301.4$\,\AA, close to the core of the bluer spectral line. All the panels are normalized to the average quiet Sun continuum. The polarization signals are shown in a logarithmic scale for better visualization. Four symbols mark the location of the profiles shown in Fig.~\ref{fig:pixel_degradation}.}
\label{fig:data}
\end{figure*}

We calculated the spectrum of the two \ion{Fe}{i} spectral lines at 630.15 and 630.25\,nm using the \textit{Stokes Inversion based on Response functions} code \citep[SIR;][]{SIR}. These spectral lines are sensitive to the temperature from the lower to the mid-photosphere. For canonical models of the solar atmosphere such as \citet{FALC}, this corresponds from $\log \tau_{500}=0$ to $\log \tau_{500} = -2$, where $\tau_{500}$ is the continuum optical depth at 500\,nm. Regarding the spectral line sensitivity of these lines, the temperature is the most important parameter as it determines the ionization state of the gas, and the populations of the relevant atomic levels, thus determining the emission and absorption properties of the plasma. They are also sensitive to line-of-sight velocity because of Doppler shifts and to the magnetic field through the Zeeman effect. The depth dependence of these physical parameters directly determines the complexity of the spectral line profiles. We carried out the synthesis, assuming local thermodynamic equilibrium (LTE), for a very fine wavelength grid spanning from 630.1 to 630.3\,nm, with a step of 5\,m\AA. Although precise modeling requires a non-LTE approach \citep{SmithaNLTE2}, differences are relatively small, and treating these spectral lines in LTE eases numerical experimentation significantly. Figure~\ref{fig:data} shows the calculated continuum intensity and the polarization close to the core of one of the two lines. 

% -------------------------------------------------------
\subsection{Spectral degradation}
% -------------------------------------------------------
The spectral resolution of a spectral discriminator (grating-based spectrograph or a Fabry-Perot filtergraph) is defined as the smallest wavelength separation $\delta \lambda$ that the instrument can distinguish. This is determined by the combination of the optical elements and the characteristics of the spectral discriminator. Spectral resolving power, a dimensionless quantity, is defined as $R=\lambda/\delta\lambda$. However, in the community, the number $R$ is often referred to as the spectral resolution so we use the same designation here. Note that the spectral fidelity of the instrument is not uniquely identified by $\delta \lambda$ or $R$, but rather by the exact shape of the so-called line spread function (LSF, also known as the spectral point spread function). That is, the recorded spectrum $I(\lambda)$ is related to the original (ideal) spectrum $I_0(\lambda)$ as:
\begin{equation}
I(\lambda) = I_0(\lambda) \star \textit{LSF} (\lambda)
\end{equation}
where $\textit{LSF}(\lambda)$ is the line spread function and $\star$ denotes the convolution in wavelength space. The LSF describes the response to a monochromatic light source: it explains how an infinitely thin spectral line (a delta function) would be at the focal plane. Furthermore, to take full advantage of the given spectral resolution, the sampling used to record the spectra should be optimal: following the Nyquist sampling criterion two pixels are used per resolution element $\delta \lambda$. The sampling we used to synthesize the data ($5$\,m\AA) is, at the given wavelength, optimal for a resolution of $R=6\times10^5$, which is several times higher than the resolution regime we are planning to investigate. Thus, we consider the spectra synthesized from the simulation to be at infinite spectral resolution compared to the spectra degraded by the LSF. 

\begin{figure*}[htp!]
\centering
\includegraphics[width=1.0\linewidth]{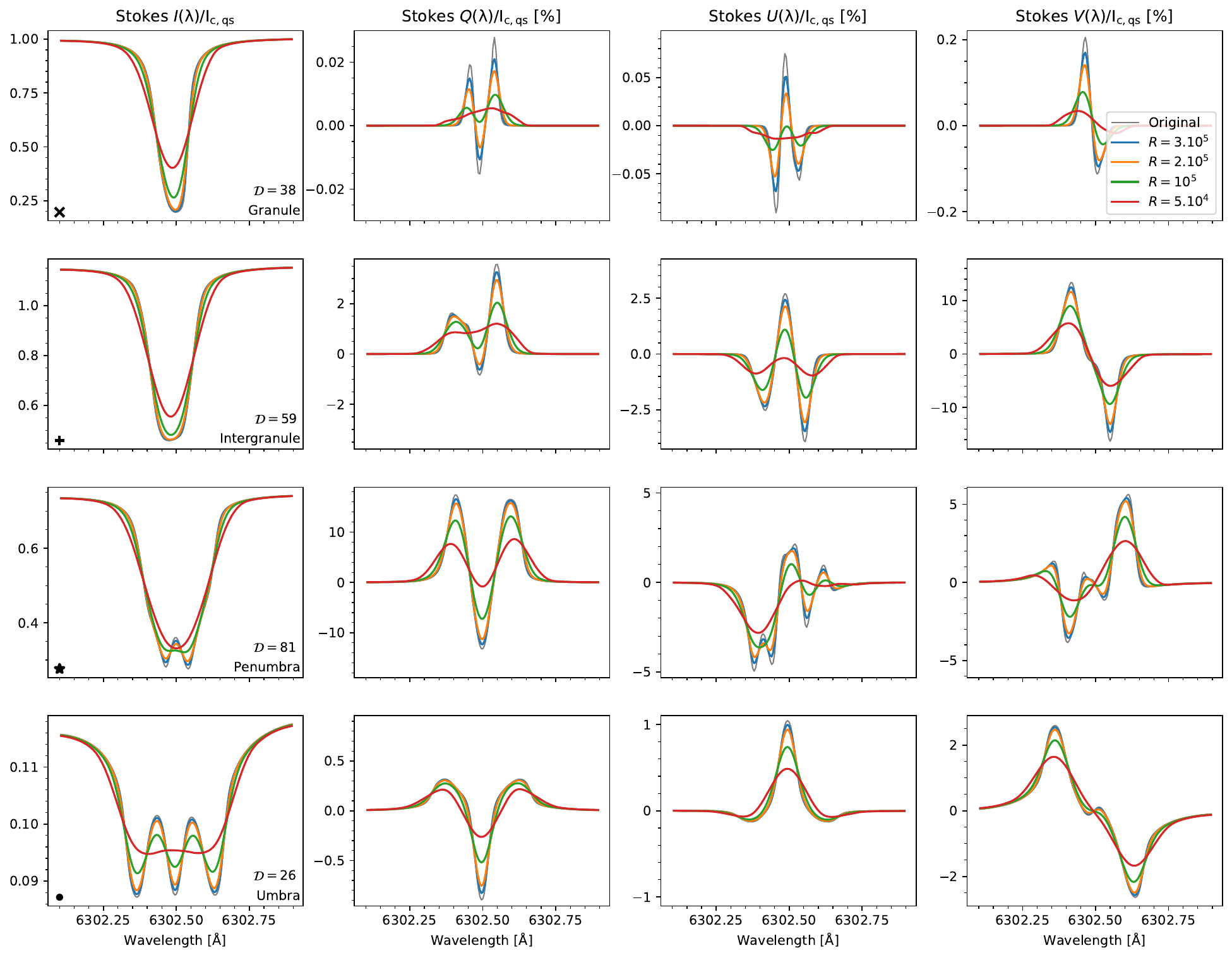}
\caption{Stokes spectra of example pixels from the simulation, under different spectral resolutions. Only one of the two \ion{Fe}{i} lines is shown, for better visibility. The location of each pixel is indicated in Fig.~\ref{fig:data} with the same symbols indicated in the lower left corner of each Stokes $I$ panel, together with the total dimensionality of that spectrum, calculated for the undegraded case.} 
\label{fig:pixel_degradation}
% This figure is made using the est_spectralres/2_PCA_data_exploration.ipynb
\end{figure*}

Depending on the properties of the instrument, the functional shape of the LSF will be different. For example, the LSF of a slit spectrograph is a $\rm sinc^2$ profile \citep[see, e.g.,][]{Casini2014JOSAAC}, which, together with other instrument imperfections, results in a final spectral profile that usually has an almost Gaussian shape \citep{Borrero2016A&A...596A...2B}. So we will consider our LSF to be Gaussian in the following analysis. It is very common to use the full-width at half maximum as the smallest wavelength separation that the instrument can distinguish, i.e. $\delta \lambda = \rm FWHM$. For the following analysis, we generate the degraded spectra by spectrally convolving the original spectra by a Gaussian corresponding to a spectral resolution of $R=10^5$. When performing spectropolarimetric inversions of these synthetic datasets, we consider spectral resolutions of $R=( 5\times 10^4, 1\times10^5, 2\times 10^5, 3\times 10^5)$ to estimate the impact of different degradations. 
% the spectral resolution of $R = 5\times 10^4$, as an example of low-resolution solar spectropolarimetry. 
Furthermore, as section \ref{sec:inversions} shows, the spectral resolution of $R=10^5$ allows a very reliable inference of atmospheric parameters, so we preferred to work on lower resolutions and sampling\ configurations. To visualize the effect of the spectral degradation, we show the Stokes profiles from four locations in the simulation (a granule, an intergranule, in the penumbra, and in the umbra) under different spectral resolutions in Fig.~\ref{fig:pixel_degradation}. As expected, the spectral degradation smears out the fine details of the Stokes profiles, especially in the polarization signals. In the context of inversions, we also study the effect that imperfect knowledge of the LSF has on the inferred atmospheric parameters. We focus on several specific scenarios because the full parameter space (number and spectral lines of interest, spectral resolution, noise level, LSF functional form, inversion configuration, etc) prevents a concise and meaningful analysis.

% -------------------------------------------------------
\section{Dimensionality analysis}\label{sec:pca}
% -------------------------------------------------------

% -------------------------------------------------------
\subsection{Dimensionality estimation}
% -------------------------------------------------------
To understand and quantify how spectral degradation affects the Stokes profiles of these specific photospheric spectral lines, we first analyze their complexity and put it in the context of the physical structure of the underlying solar atmosphere. More complex profiles are expected to be harder to model with spectropolarimetric inversion techniques but potentially hold more information. Very complicated or unusual profiles will point to interesting atmospheric structures with potential for scientific discovery. 

\begin{figure}[htp!]
\centering
\includegraphics[width=\linewidth,trim={0cm 0cm 0cm 0cm},clip]{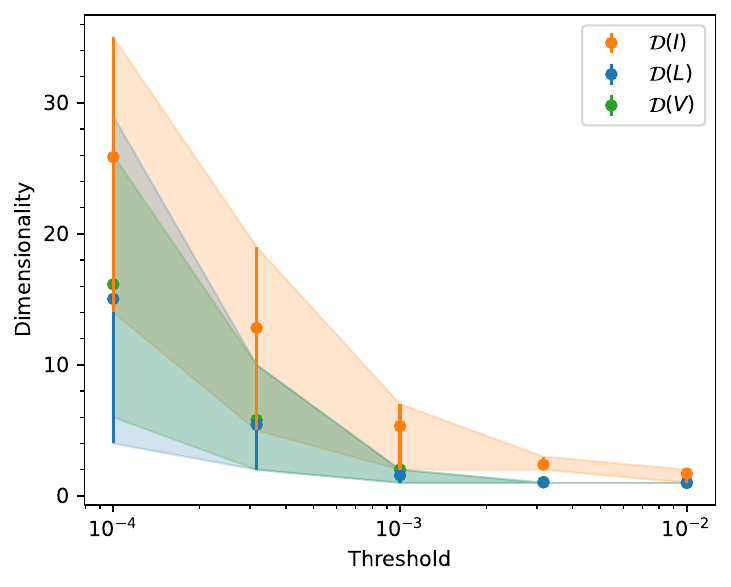}
\caption{Spectra dimensionality calculated by PCA based on different thresholds for the reconstruction of the $5\times10^5$ Stokes profiles from the MURaM snapshot. This is estimated on the spectra containing the two \ion{Fe}{i} spectral lines at full spectral resolution. The filled circles are the average dimensionality and the color bands (and error bars) show the 16th and 84th percentiles of the distribution.}
\label{fig:dim_th}
\end{figure}

Motivated by the work of \cite{Asensio2007ApJ...660.1690A}, we quantify the complexity of the Stokes profiles by calculating their dimensionality using Principal Components Analysis \citep[PCA, e.g.][]{num_rec} which allows us to decompose a Stokes profile into a series of orthogonal components ordered according to the amount of variance they explain. In general, the number of complex profiles is a small fraction of the total number of profiles, so the ordering of the PCA components according to the variance is compatible with an ordering according to their complexity. Thus we use the number of components required to reproduce a profile as a measure of its dimensionality \citep{Marian2008A&A...486..637M}. To compute the dimensionality $\mathcal{D}$ of each Stokes spectrum we first create the set of basis vectors, separately for intensity, linear, and circular polarization, using the entire field of view. For simplicity, from now on the linear polarization (defined as $L=\sqrt{Q^2+U^2}$) will be analyzed instead of the Stokes $Q$ and $U$ separately. We define the dimensionality of each spectrum as the number of components needed to reproduce the profile $S$ with a standard deviation lower than a given threshold $\sigma$  \citep{Borrero2016A&A...596A...2B}. That is, 
\begin{equation} 
\mathcal{D}(S) = \min_{N} \left\{ N \, | \, \sqrt{ \left( \sum_{i=0}^N c_i \mathcal{V}_i - S \right)^2}/N_w < \sigma \right\} ,
\label{eq:pca}\end{equation}
% \mathcal{S}
where $i$ enumerates the order of the basis vectors, $c_i$ are the coefficients in the PCA decomposition, $\mathcal{V}_i$ the eigenvectors of the basis, and $N_w$ the number of wavelength points in the Stokes parameter $S$. Figure~\ref{fig:dim_th} shows the average dimensionality of each Stokes parameter on different thresholds for the 524288 ($256\times2048$) Stokes profiles of the MURaM snapshot. In this figure, the dimensionality of the Stokes parameters rises very rapidly as the threshold decreases (note the logarithmic scale in the horizontal axis). For a threshold between $10^{-2}$ and $10^{-3}$, the spectra can be explained mainly with 2--5 components. The very small-scale features are only identified when the threshold is lowered to $10^{-4}$, which suggests that the Stokes parameters are usually simple to explain and the substructure of the profiles has a smaller amplitude. Our choice of threshold is driven by typical photon noise found in spatially resolved solar spectropolarimetric observations. In the following, the dimensionality has been calculated using $\sigma=10^{-4}$ for all the Stokes parameters. We have verified that the overall results are not affected by the specific choice. 

Note that our threshold is defined in an absolute way. So, Stokes profiles with very weak polarization signals, or weak spectral features will be classified as low-dimensional even though they might appear very complicated. This is because the criterion from the Eq.\,\ref{eq:pca} is satisfied already for a small $N$, because a few first basis vectors are enough to reproduce the profile closely to the threshold. Said differently, our approach classifies profiles according to the dimensionality we can detect with our limited signal-to-noise ratio. Alternative approach would be to use Eq.\,\ref{eq:chi2} with the threshold defined in a relative way (for example, fraction of the maximum polarization signal) and thus characterize the complexity of the low-amplitude signals. This approach, would, on the other side, identify complex profiles that can not be detected in actual observations, due to the photon noise.

% -------------------------------------------------------
\subsection{Spatial distribution of dimensionality}
% -------------------------------------------------------

\begin{figure*}[htp!]
\centering
\includegraphics[width=1.0\textwidth]{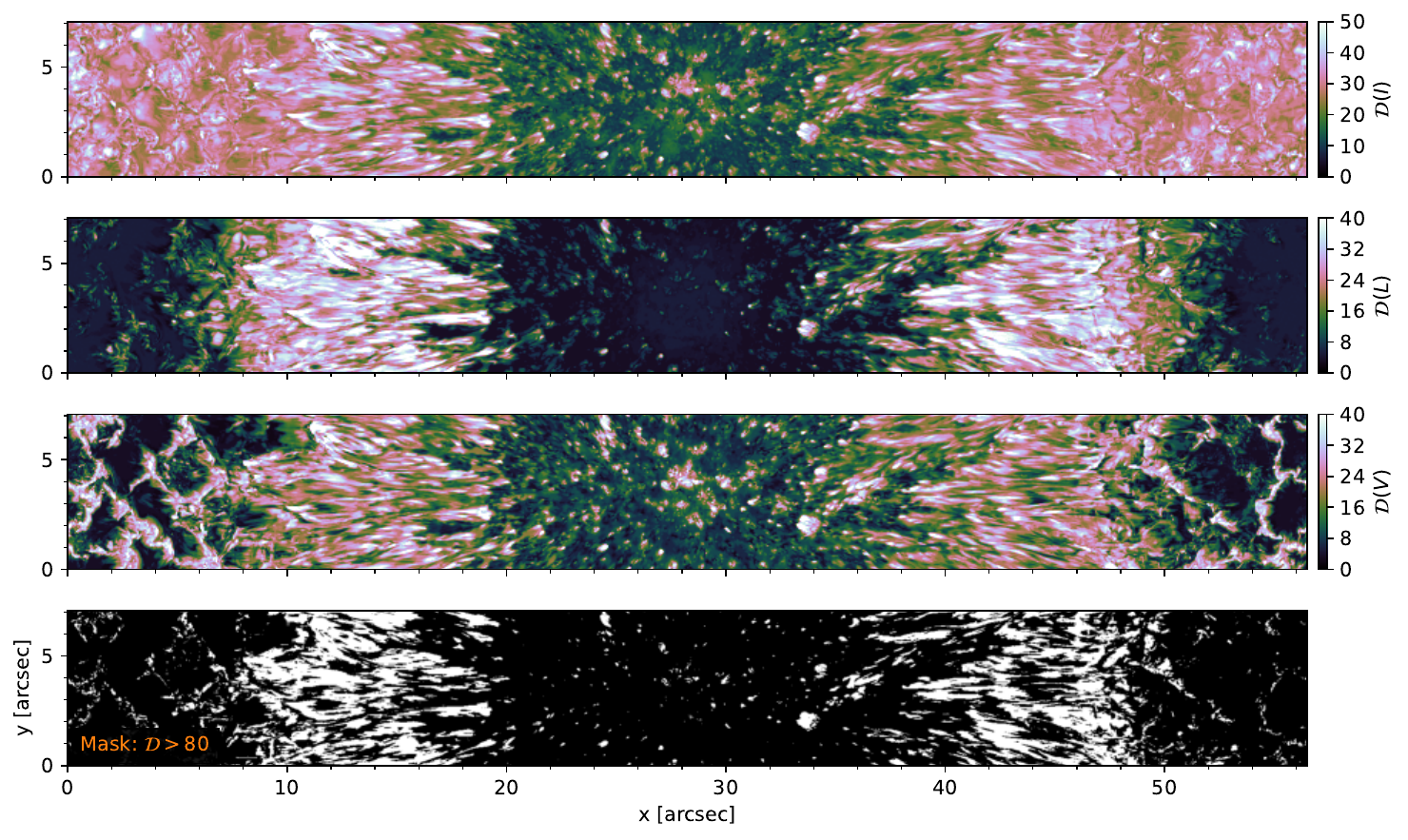}
\caption{Spatial distribution of the dimensionality of Stokes $I$ (top panel), linear polarization (second panel), and Stokes $V$ (third panel), defined as the number of PCA components needed to reproduce such signals under a threshold of $\sigma=10^{-4}$. The bottom panel shows a binary mask where white pixels mark the spectra whose total dimensionality $\mathcal{D}(I)$ + $\mathcal{D}(L)$ + $\mathcal{D}(V)$ is larger than 80.}
% This figure is made using the est_spectralres/2_PCA_data_exploration.ipynb
\label{fig:dimensionality}
\end{figure*}

Figure~\ref{fig:dimensionality} shows the spatial distribution of the dimensionality of the Stokes profiles over the field of view. The top panel shows the dimensionality of Stokes $I$. The distribution is very homogeneous, having an average value of 30 in the weakly or intermediate magnetized regions and decreasing to 15 in the umbra. Only very few locations (1.2\% of the FoV) need more than 50 components, where this value can go up to 110. The second row shows the spatial distribution for the linear polarization signals and their dimensionality resembles very closely their amplitude (see the second and third panels of Fig.~\ref{fig:data}). A high number of PCA components is needed to reproduce the signals in the penumbra because of its strong and complicated magnetic fields and the presence of velocity field gradients. Outside the penumbra, the signals might appear to be complex but their amplitudes are well below the specified threshold. Again, this is consequence of our definition of dimensionality.

The dimensionality of Stokes $V$ (third panel) presents high values in the intergranular lanes and the core of the penumbral filaments. On the contrary, the profiles that emerge from the umbra have a very low dimensionality. This can be understood as a consequence of the homogeneous properties of the solar atmosphere in those particular regions, i.e., the temperature, the velocity, and the magnetic field also show a decrease in the dimensionality (see Fig.~\ref{fig:dimensionalitysimulation}). Some umbral dots appear to have an increased dimensionality (in particular in Stokes $V$) which is not well visible in the dimensionality of the physical quantities. This behavior can be interpreted as an indication of the non-linear radiative transfer process that generates the observed Stokes profiles. 

Lastly, to compare the results of the degradation in pixels with different complexity, we calculated a binary mask (shown at the bottom panel of the same figure) of the spectra whose total dimensionality (sum of individual dimensionalities, i.e., $\mathcal{D}(I)$ + $\mathcal{D}(L)$ + $\mathcal{D}(V)$) is larger than a specific threshold. This will be used later to distinguish \textit{simple} from \textit{complex} profiles. A value of 80 is a good compromise to capture an important percentage ($\sim 20 \%$) of the profiles that present small-scale features in the region. To understand the relation between the estimated dimensionality and the typical shapes of the Stokes profiles, Fig.~\ref{fig:pixel_degradation}  also displays the total dimensionality in the bottom right corner of each of the extracted pixels.

% -------------------------------------------------------
\subsection{Influence of limited spectral resolution on the dimensionality}
% -------------------------------------------------------

\begin{figure*}[htp!]
\centering
\includegraphics[width=1\linewidth]{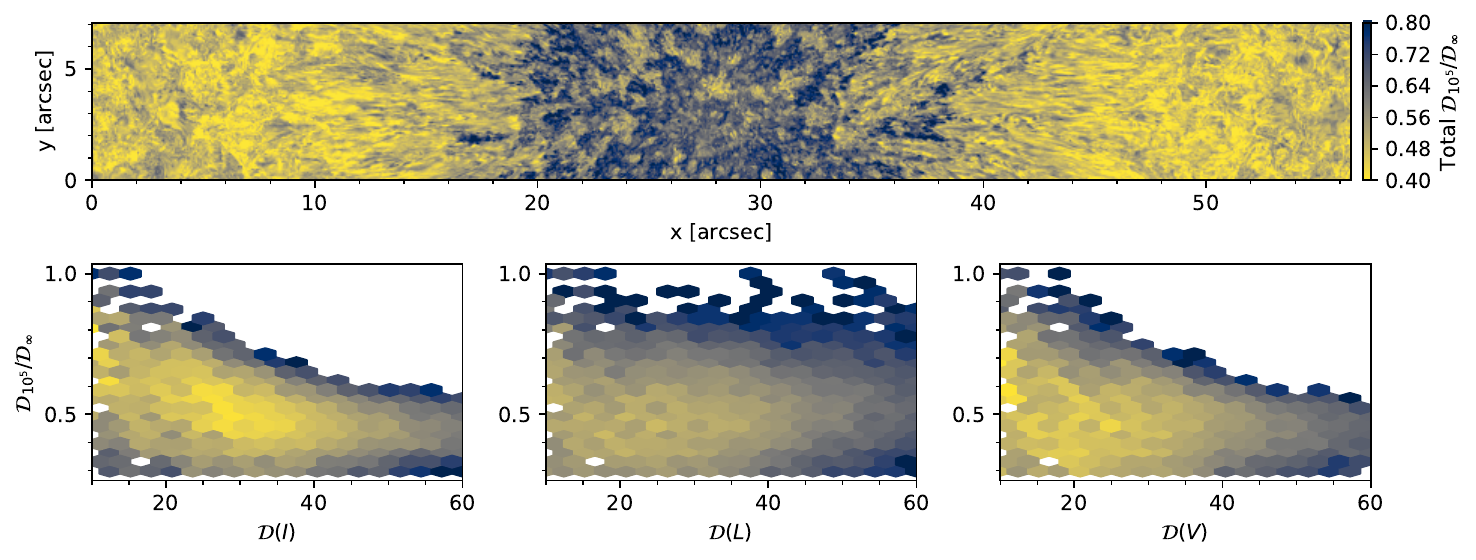}
\caption{Difference in the dimensionality of the original and degraded spectra emerging from the simulation. Upper panel: Ratio of the dimensionality calculated from the original spectra emerged from the simulation and after degrading it to a spectral resolution of $R=10^5$ (smaller values are regions where the degraded spectra are more affected). Bottom row: 2D histograms of the ratio of the dimensionality for each Stokes parameter.}
\label{fig:ratio}
\end{figure*}

To quantify the impact of the spectral degradation on the dimensionality of the Stokes profiles, we degraded the original synthetic profiles to a spectral resolution of $R=10^5$ and calculated the dimensionality with the same procedure (i.e., using Eq.~\ref{eq:pca}). Figure~\ref{fig:ratio} shows the ratio between the dimensionality before (formulated as $\mathcal{D}_\infty$) and after the degradation, using the same threshold $\sigma$, for the total dimensionality (upper panel), and for each individual Stokes parameter (bottom row).

As expected, the degradation decreases the dimensionality of the data, i.e., the ratio is always lower than 1 across the region. The spatial distribution of the upper panel shows that the umbra is less affected by the degradation. Regarding individual Stokes contributions, there is a trend where the dimensionality tends to decrease by half for pixels with high dimensionality. This is more visible in Stokes $I$ and $V$ than in the linear polarization. We would expect that pixels with lower dimensionality should not change much because the degradation should affect mainly profiles where many components are needed to reproduce small-scale features. However, the 2D histograms of each Stokes parameter show an average value lower than unity also at low $\mathcal{D}$ values. The fact that each dataset is calculated using a new PCA basis, could explain that the new degraded data needs fewer eigenvectors to efficiently explain the observations.

In summary, the PCA decomposition can be used to evaluate the complexity of the profiles. The largest change of dimensionality is found on the profiles with the highest original dimensionality, i.e., the spectra emerging from the penumbra and intergranular lanes.

\begin{figure*}[htp!]
\centering
\includegraphics[width=1\linewidth]{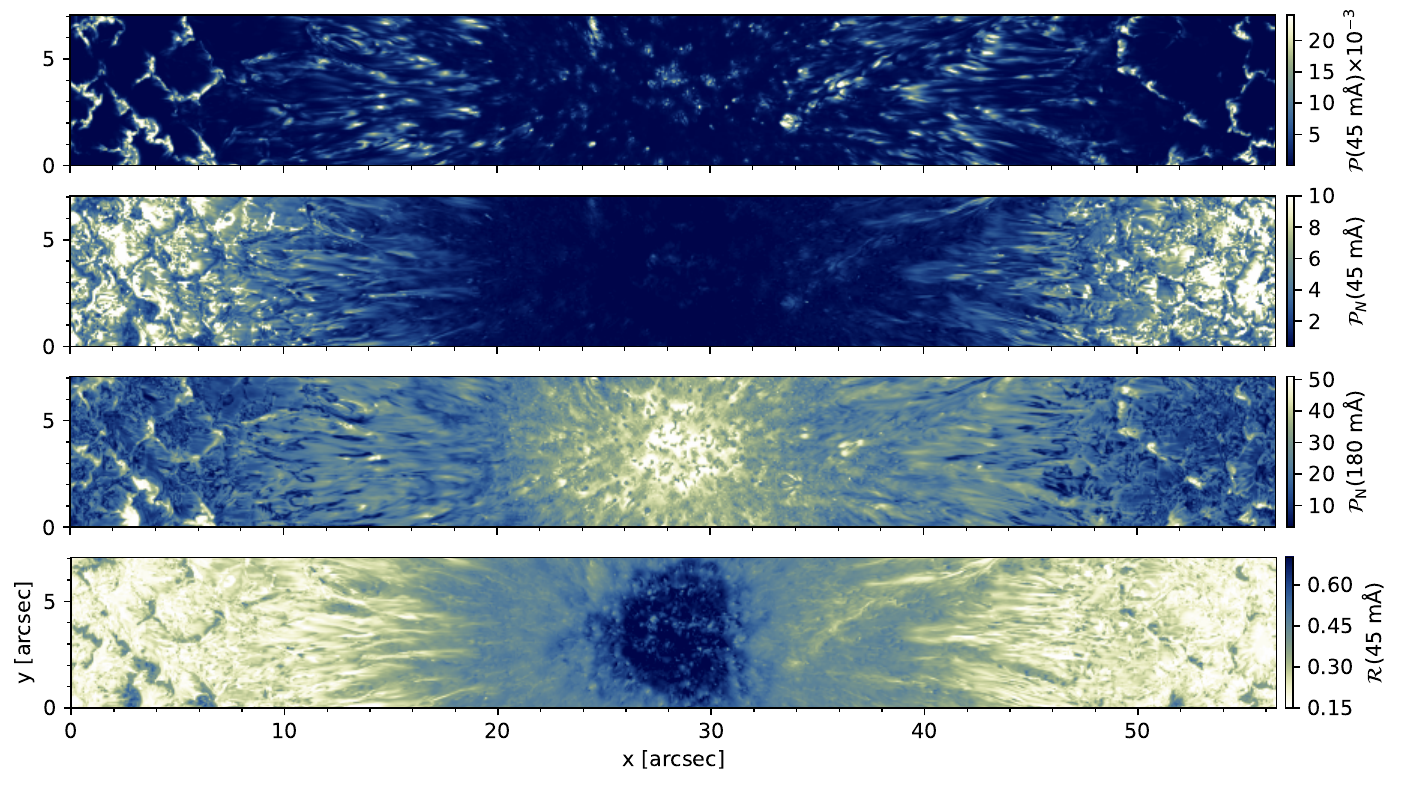}
\caption{Summary of the wavelet analysis of the Stokes $V$ profiles. The first panel shows the power contained in the wavelet decomposition at scales of 45~m\AA. The second and third panels show the power contained in the normalized spectra at scales of 45 and 180~m\AA, respectively. The fourth panel shows the ratio of the power before and after the degradation at a scale of 45~m\AA.}
\label{fig:wavelet_summary}
\end{figure*}

% -------------------------------------------------------
\section{Wavelet analysis}\label{sec:wavelet}
% -------------------------------------------------------

% -------------------------------------------------------
\subsection{Wavelet decomposition}
% -------------------------------------------------------
It is expected that a limited spectral resolution will suppress the small-scale spectral features. To quantify this effect, we analyzed the power contained in different spectral scales in the Stokes profiles. We chose the wavelet decomposition technique instead of Fourier decomposition because the signals (the Stokes profiles) are confined in the original domain (wavelength range). Fourier decomposition is made in sinusoidal signals defined across the whole domain, while in wavelet analysis the base signal (i.e., \textit{the mother wavelet}) has a confined functional form and can be shifted and scaled to cover the domain. Given the similarity with the Stokes profiles, the DOG wavelet (Derivative of Gaussian) is chosen as the mother wavelet. We define the \textit{spectral scale} of our profiles as half of the wavelet period. The decomposition is performed using the \texttt{pycwt}\footnote{\url{https://github.com/regeirk/pycwt}} python package. The wavelet coefficients are then used to compute the wavelet power spectrum, defined as the square of the absolute value of the wavelet coefficients. Note that, depending on the mother wavelet, the power spectrum can be slightly different, so we should treat these results as a representative behavior and not as a precise description of the scales in the spectra.

% -------------------------------------------------------
\subsection{Degradation of spectral scales}
% -------------------------------------------------------

In the following analysis, we focus on Stokes $V$ signals, but the results for other Stokes components are similar. The first panel of Fig.~\ref{fig:wavelet_summary} shows the power $\mathcal{P}$ contained in the wavelet decomposition at scales of 45~m\AA\ for the Stokes $V$ signals. We chose this scale as it is slightly below the FWHM that corresponds to the $10^5$ resolution at 630\,nm, and we thus expect it to be substantially influenced by the finite spectral resolution. However, the power at any other scale presents a very similar distribution because the power scales with the square of the signal amplitude, and the pixels with the strongest signals have the most power. This is, in a way, similar to the results of the previous section where we found that the pixels with larger polarization amplitudes show higher dimensionality. 

To focus on the spectral shape and not on the amplitude, we normalized each pixel to its maximum. This way we can analyze the spectral scales present in the profiles. The power contained in the normalized spectra $\mathcal{P}_{\rm N}$ at two different wavelength scales is shown in the second and third panels of the same figure. The smallest scales (second panel) are found within the granules, while the larger scales (third panel) are found in intergranules, penumbra, and umbra. This is expected because the magnetic field broadens the Stokes $V$ profiles. From this analysis we can conclude that although there are many pixels with small spectral scales, they will be impossible to detect under a specific signal-to-noise ratio, confirming the results from the PCA analysis. 

When the spectra are degraded to $R=10^5$, the smallest scales (similar to or shorter than the width of the LSF) are the most affected by the degradation and larger scales are mostly unaffected. To show this, we calculate the ratio $\mathcal{R}$ of the power before and after the degradation (without normalizing the data). This is shown in the fourth panel of Fig.~\ref{fig:wavelet_summary} only for the small scales (45~m\AA). The ratio is always lower than 1, with very low values not only within the granules but also in many locations inside the penumbra. A good example of how the spectral degradation smears out these smallest scales in those regions is shown in Fig.~\ref{fig:pixel_degradation}.

% -------------------------------------------------------
\section{Spectropolarimetric Inversions}\label{sec:inversions}
% -------------------------------------------------------

Probably the most relevant test of information loss is to perform an end-to-end study of the inference process \citep[see e.g.][]{delaCruz2012A&A...543A..34D, Millic2019A&A...630A.133M}. We use the synthetic data as our observables and the inferred physical parameters are compared to the original stratification, i.e. to the \textit{true} solution, thus investigating how spectral degradation affects the inference process. To ensure the absence of biases related to model atoms, opacity packages, and specific numerical schemes, we once again employ the SIR code for the inversion. This approach is preferred as it helps avoid discrepancies that frequently arise when using different inversion codes. We have implemented our own MPI parallelized version to speed up the inversions (see Appendix~\ref{app:app2} for more information). In the following, we present the results of the inversions considering different levels of spectral resolutions, binning, and photon noise. We are motivated by the instrument requirements for the 4-m class telescopes \citep{DKIST2020, QuinteroNoda2022}, so we do not perform any spatial degradation.

% -------------------------------------------------------
\subsection{Inversion complexity}\label{invconf}
% -------------------------------------------------------

Inversion is an optimization process that tries to find the model atmosphere (i.e. set of depth-dependent physical parameters) that best reproduces the observed Stokes profiles. In SIR and other stratified inversion codes such as SNAPI \citep{SNAPI}, STiC \citep{STIC} or FIRTEZ \citep{PastorYabar2019A&A}, the model atmosphere is parametrized at some locations in the optical depth scale (called \textit{nodes}) and the remaining part of the atmosphere is obtained by interpolating a perturbation at the nodes (SIR, FIRTEZ) or by interpolating values at the nodes themselves (SNAPI, STiC). The nodes are the free parameters of our model and the number of nodes represents a measure of the complexity of the model. Inversion is generally an ill-posed problem which means that many combinations of parameters yield equally good fits to the observations (i.e. the parameters are degenerate). This degeneracy increases with the complexity of our model. Using many nodes can result in overfitting, where the inferred atmosphere presents oscillatory or unrealistic solutions. On the other hand, using too few nodes can lead to underfitting, where the model cannot reproduce the observed profiles.

To understand the interaction between the complexity of the spectra and the complexity of the model, we use three configurations with varying numbers of nodes: a \textit{minimal} configuration with a small number of nodes that provides a good estimation (configuration~1), a \textit{robust} configuration that captures most of the features (configuration~2), and a \textit{sophisticated} configuration that provides a very good fit to the data (configuration~3). The parameters that are allowed to change are the temperature (T), the line-of-sight velocity ($v_{\rm LOS}$), the magnetic field strength ($\rm B$), the inclination of the magnetic field with respect to the line of sight ($\Theta_B$), and the azimuth of the magnetic field in the plane perpendicular to the line of sight ($\Phi_B$). All the configurations have three cycles, i.e., we run three times the inversion successively increasing the number of nodes until reaching the final number of nodes per physical parameter (see Table\,\ref{tab:hyperparameters} for details). Additionally, to achieve as good inversion as possible, the results of the robust configuration are used as an input for the complex configuration. 
% Note that SIR uses no additional regularization, so the complexity of our models is determined purely by the number of nodes used.

\begin{table}[!ht]
    \centering
    \begin{tabular}{c|c|c|c}
        Parameters& Config 1 & Config 2 & Config 3 \\
        \hline
        \hline
        T & 2, 3, 4 & 3, 5, 7 & 4, 7, 10 \\
        $v_{\rm LOS}$ & 1, 2, 3 & 2, 4, 7 & 3, 4, 10 \\
        B & 1, 2, 3 & 1, 2, 4 & 3, 4, 10 \\
        $\Theta_B$ & 1, 1, 2 & 1, 2, 4 & 2, 3, 10 \\
        $\Phi_B$ & 1, 1, 2 & 1, 2, 4 & 2, 3, 4 \\
    \end{tabular}
    \caption{Hyperparameters used in the inversions: nodes configuration for the physical parameters depending on the scheme.}
    \label{tab:hyperparameters}
\end{table}
    
% -------------------------------------------------------
\subsection{Instrumental effects}
% -------------------------------------------------------

To quantify the impact of the spectral resolution on the inferred physical parameters we created and inverted datasets where different instrumental effects were applied: {\sc i}) the original spectra from the simulation, {\sc ii}) the spectrally degraded spectra, {\sc iii}) the spectrally degraded spectra with noise, and {\sc iv}) the spectrally degraded spectra with noise, spectrally resampled according to the Nyquist-Shannon theorem for the corresponding spectral resolution. At the original sampling of 5\,m\AA, we estimate the number of photons per bin in the following way \citep[similar to][]{Tino2019A&A}:
\begin{equation}
    N = \frac{2c^3}{\lambda^4 (\exp(hc/\lambda k T) - 1)} \times \frac{D^2 \pi}{4 d^2} \times \Delta x^2 \times \Delta t \times \Delta\lambda \ \times \eta.
\end{equation}
where the first factor on the right-hand side is the number of photons emitted per unit surface in unit solid angle per unit time per unit wavelength, the second is the solid angle spanned by the telescope, the third is the surface on the Sun corresponding to one pixel, fourth is exposure time, the fifth is the size of wavelength bin and the sixth is the efficiency of the telescope-spectrograph system. We took $T=5700\,$K, $\lambda = 630$\,nm, $D=4\,$m, $\Delta x=14$\,km (half of the diffraction limit for corresponding $D$ calculated as 1.22$\lambda/D$), $\Delta t = 1\,$s, $d=1.5\cdot10^{11}$m (distance Earth - Sun) and $\Delta\lambda=5$\,m\AA. For efficiency, we took a very conservative $\eta=0.03$, which is on the lower end for ground-based observatories (in Hinode/SOT/SP is estimated to have an efficiency of 0.25). This brings us to a signal-to-noise (SNR) ratio of around 350 in the Stokes $I$ continuum, and, assuming the optimal demodulation ($1/\sqrt{3}$), approximately SNR of 200 in other Stokes components. Following this estimate, we applied a noise of $5 \times 10^{-3}$ in units of the continuum intensity to all four Stokes components, scaling the noise with the square root of the Stokes $I$ continuum in each pixel. 

For the resampled data, the Nyquist-Shannon theorem states that the sampling should be at least 2 pixels per resolution element. Under this spectral resolution, the spectral lines have been sampled to a pixel size of 30 m\AA. In that case, the noise level per spectral bin is decreased by more than a factor of 2, to $2 \times 10^{-3}$ in units of the continuum intensity. To contextualize these values, the spectral resolution of the Hinode/SP instrument is around $R=2\times 10^5$ with $21.5$~m\AA\ sampling \citep{Lites_SP}. Our noise levels are higher than a typical Hinode/SP observation due to shorter exposure and a low estimate of $\eta$. Better noise levels can be achieved with longer exposure times, but we have preferred to be conservative in this step. Other reference values of spectral resolution at this wavelength are added for comparison: about 105\,000 for VTF/DKIST \citep{VTF2014}, 180\,000 ViSP/DKIST \citep{deWijn2022}, 170\,000 SPINOR/DST \citep{SocasNavarro2006SoPh}, about 115\,000 for CRISP/SST \citep{CRISP} and about 200\,000 for TRIPPEL/SST \citep{Kiselman2011A&A}.

% GRIS 190 000 in the 10830.
All four sets of the data -- {\sc i}) original, {\sc ii}) convolved, {\sc iii}) convolved and noised and {\sc iv}) convolved, noised, and resampled -- have been inverted using each of the three configurations described in Sec.~\ref{invconf}.

\begin{figure}[t!]
\centering
\includegraphics[width=\linewidth]{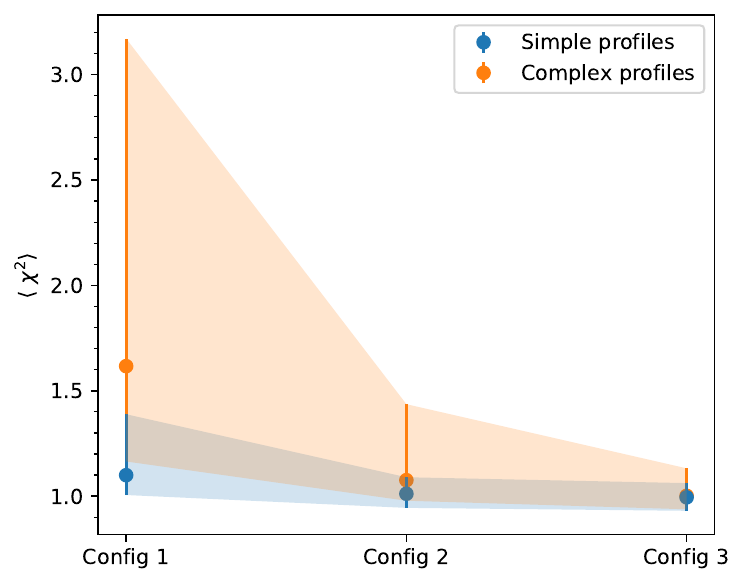}
\caption{Quality of the inversions for the different node configurations of increasing complexity (see Table~\ref{tab:hyperparameters}). The filled circles are the average $\chi^2$ (Eq.~\ref{eq:chi2}) calculated for profiles whose total dimensionality is smaller (\textit{simple}) or larger (\textit{complex}) than 80 (see binary mask in Fig.~\ref{fig:dimensionality}). The color bands (and error bars) show the 16th and 84th percentiles of the each distribution.}
\label{fig:chi2_configs}
\end{figure}

\begin{figure*}[t!]
\centering
\includegraphics[width=1\linewidth]{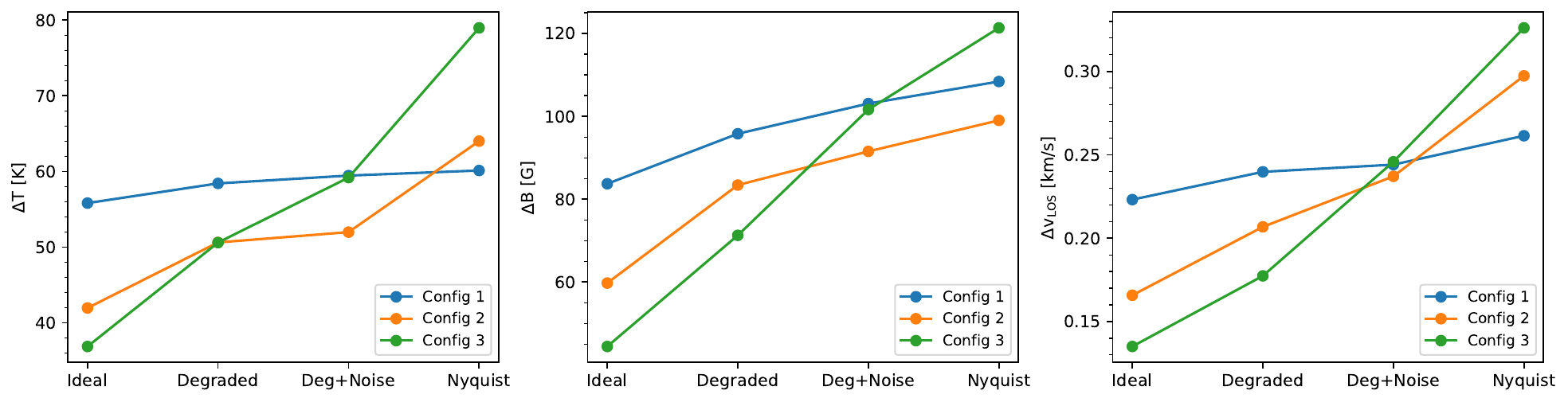}
\caption{Average error between the inferred and the original values from the simulation in the range $\log \tau_{500}=[-0.5,-1.5]$ for different scenarios and different inversion configurations. The error is measured as the standard deviation. The left panel shows the error in temperature, the middle magnetic field strength, and the right panel velocity. In each panel from left to right: inversions using the original spectra, degraded spectra, degraded spectra with noise, and degraded spectra with noise sampled according to the Nyquist-Shannon theorem at the corresponding spectral resolution.}
% rouppe: Note that original is called "Ideal" on the x-axis labels
\label{fig:mean_error}
\end{figure*}

\begin{figure*}[t!]
\centering
\includegraphics[width=1\linewidth]{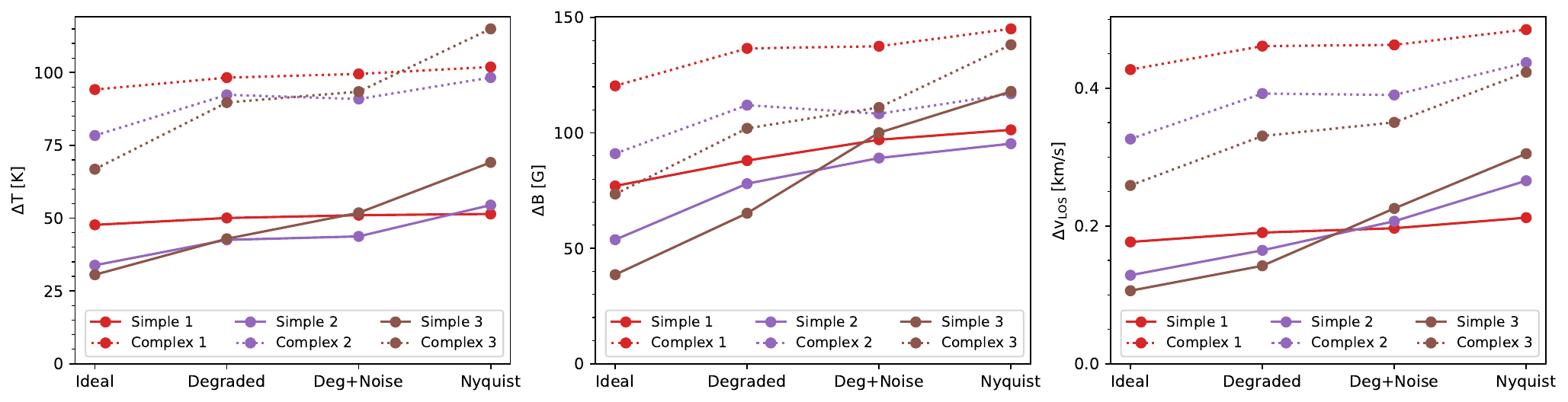}
\caption{Same as Fig.~\ref{fig:mean_error} but for the simple (solid lines) and complex (dotted lines) Stokes profiles.}
\label{fig:simple_complex}
\end{figure*}

% -------------------------------------------------------
\subsection{Inversion results}
% -------------------------------------------------------

Before analyzing the inferred physical quantities, we studied the performance of each node configuration when reproducing the Stokes profiles. A measure of the quality of the fit,
\begin{equation}
    \chi^2 = \frac{1}{4 N_w } \sum_{w}^{N_w} \sum_{i}^4 \frac{\left(S_{i,w}^{\rm inv} - S_{i,w}^{\rm obs} \right)^2}{\sigma_i^2}
    \label{eq:chi2}
\end{equation}
is calculated for each pixel, where $N_w$ is the number of wavelength points, $S_i^{\rm inv}$ and $S_i^{\rm obs}$ are the inverted and observed Stokes profiles, respectively, and $\sigma_i$ is the noise level. We consider that the inversion is able to reproduce the observed profiles when $\chi^2 \approx 1$. The average quality of each configuration $\langle\chi^2\rangle$ is shown in Fig.~\ref{fig:chi2_configs} for the noisy degraded case. In this figure, we have also separated the simple and complex profiles according to the mask defined in Fig.~\ref{fig:dimensionality}. This figure shows that the simple profiles are well reproduced by all configurations. On the other hand, the complex profiles are better reproduced when increasing the complexity of the configuration. This ensures that worse retrievals using the sophisticated configuration are not due to a problem when reproducing the profiles but due to the over-complicated solutions.

The results of the inversions of the four datasets are displayed in Fig.~\ref{fig:mean_error}. This figure depicts the standard deviation between the inferred and the true values for the temperature, magnetic field strength, and line-of-sight velocity across the FoV in the range of heights with larger sensitivity (between $\log \tau_{500}=-0.5$ and $\log \tau_{500} = -1.5$). This quantity can be understood as an average error or discrepancy between the inferred and true values. Reducing the range of optical depths to a single point in height (at $\log \tau_{500} = -1$) has almost no impact on the spatially averaged errors. The results are presented for the different configurations and the different scenarios. 

For the original scenario where no degradation has been applied, the complex configuration (indicated by the green solid line) achieves superior accuracy, exhibiting average errors of approximately 38~K for temperature, 45~G for magnetic field strength, and 0.14~km/s for line-of-sight velocity. However, as soon as we degrade the data, the error of each configuration increases. This effect is particularly significant in the most complex configuration. This is because the complex configuration is more prone to overfitting, and the data degradation makes the inversion process more ill-posed (i.e., more solutions can reproduce the observed Stokes profiles). The presence of noise significantly influences the accuracy of inversion results. The complex configuration becomes as bad as the simple configuration, while the robust configuration provides the best results. 

Lastly, when the spectra are resampled, the robust configuration also provides worse results than the simple configuration (most of the time). While the noise level per spectral bin is lower than in the degraded+noise case, spectral sampling makes the recovery of the physical parameters more difficult. In fact, the typical error at this point is of the order of twice the error of the ideal case. In summary, the best-performing configuration depends on the amount of information present in the data. This also shows, that for this particular case, the amount of information we lose when resampling the data is larger than what we gain in signal-to-noise ratio. It is important to note, however, that sparser wavelength sampling allows observations with a wider wavelength range, which could increase the stratified information by combining more spectral lines \citep{Tino2019A&A}.

We repeated the same analysis to quantify how different inversion configurations perform on spectra of different complexity. For that, we have used a mask defined in Sect.~\ref{sec:pca}, and the results when being split into these two groups are shown in  Fig.~\ref{fig:simple_complex}. One might think that complex profiles could produce distinctive imprints from atmospheric conditions that would make the inverse process better constrained, but in fact, complex profiles tend to have larger errors. This could be due to forward modeling and loss of information as part of radiative transfer, or in the inversion process due to e.g. further degeneracy in modeling, or the validity of the hydrostatic equilibrium hypothesis. Another result is that the distance between these two groups is smaller for the magnetic field strength than for the temperature and line-of-sight velocity. This result can be understood if the magnetic field strength in simple and complex profiles is not as different, but the temperature and line-of-sight velocity make the profiles look much more complex.

\begin{figure}[t!]
\centering
\includegraphics[width=\linewidth,trim={0cm 0cm 0cm 0cm},clip]{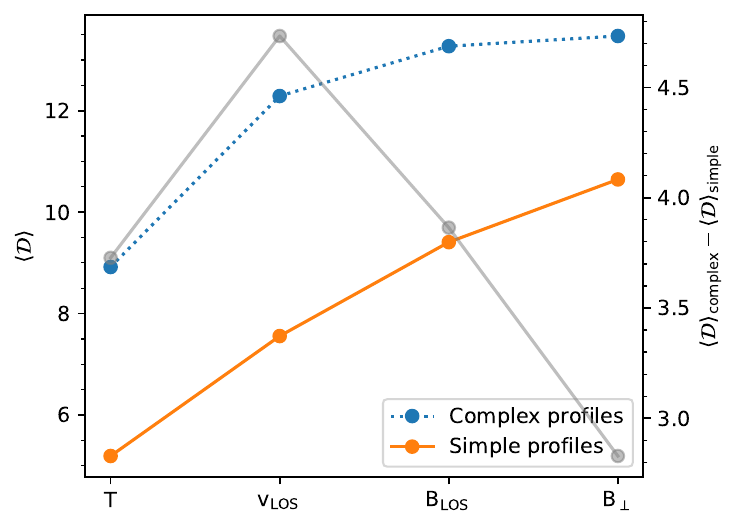}
\caption{Average dimensionality of the physical quantities of our simulation from stratifications belonging to simple and complex profiles. The gray solid line represents the difference between the average dimensionality of simple and complex profiles.}
\label{fig:model_complexity}
% This figure is made using the est_spectralres/2_PCA_data_exploration.ipynb
\end{figure}

To further investigate the impact of the complexity of the depth stratification of the physical quantities on the inversion results, we have calculated the dimensionality of the physical quantities of our simulation (see Appendix\,\ref{app:app1}). Later, we calculated the average dimensionality of the physical quantities for the simple and complex profiles. These results are shown in Fig.~\ref{fig:model_complexity}. As expected, the complex profiles come from regions with a higher dimensionality in the physical quantities. In particular, if we look at the difference between the simple and complex profiles (gray solid line), we can see that the line-of-sight velocity is the physical quantity that shows the largest difference between the simple and complex profiles. In conclusion, the complexity of the Stokes profiles is mostly controlled by the gradients in the line-of-sight velocity (according to our simulation).

% -------------------------------------------------------
\subsection{Additional tests}
% -------------------------------------------------------

Our tests show visible effects of spectral resolution, binning, and noise on the inferred parameters. Still, the agreement between the original atmosphere and inferred parameters with varying degrees of instrumental effects is excellent: below 100\,K in temperature, up to 100~G in the magnetic field, and up to 0.3~km/s in LOS velocity. In reality, these disagreements are likely to be shadowed by various systematics: imperfect knowledge of atomic parameters, inadequacy of assumption of LTE, or not detailed enough knowledge of instrumental effects. It is not straightforward to perform an end-to-end study taking into account the effects one does not know about. Nevertheless, to take our inversion scheme to the brink of applicability we performed four more inversion tests: {\sc i}) using only one spectral line from the line pair (630.25 nm), {\sc ii}) using a low-spectral resolution case with $R=5 \times 10^4$, {\sc iii}) treating spectral resolution as unknown and fitting the width of the LSF as a free parameter (through the macroturbulent velocity parameter), and {\sc iv}) degrading the spectra with a different LSF (${\rm sinc}^2$ function), but using a Gaussian during the inversion. These tests aim to mimic a situation in which our information space is limited and/or we do not know our instrumental effects well enough. 

\begin{figure}[t!]
\centering
\includegraphics[width=\linewidth,trim={0cm 0cm 0cm 0cm},clip]{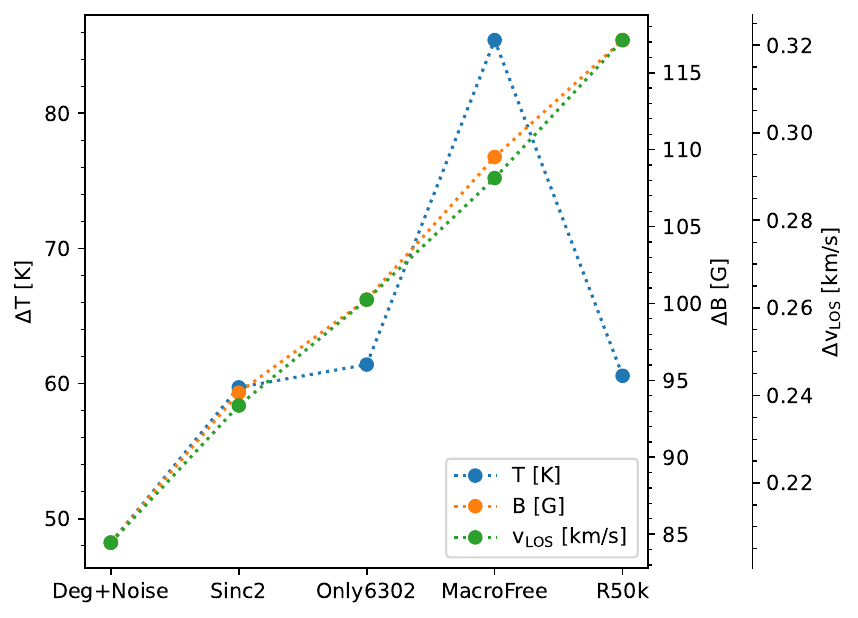}
\caption{Average error in the temperature, magnetic field strength, and line-of-sight velocity for the case where a $\rm sinc^2$ LSF is modeled with a Gaussian, when only one spectral line is used (630.15 nm), when the macroturbulent velocity parameter is let to vary in every pixel, or when having a much lower spectral resolution ($R=5\times10^4$) compared to the noisy degraded case ($R=10^5$).}
\label{fig:onlyLine_macroturbulence}
% This figure is made using the est_spectralres/4_AnalyzingInversions_summary_noparams.ipynb
\end{figure}

The results of these runs are shown in Fig.~\ref{fig:onlyLine_macroturbulence}. Starting with the case with the lower impact, if the ${\rm sinc}^2$ LSF is approximated by a Gaussian the errors increase by about 20\% in temperature, 12\% in magnetic field strength, and 20\% in line-of-sight velocity. They might seem large percentages but in absolute units, they represent still low errors, indicating that the inversion process is not very affected by the choice of the LSF. 
% (60-50)/50*100 = 20%
% (95-85)/85*100 = 11.8%
% (0.24-0.20)/0.20*100 = 20%
Using only one spectral line the error increases in the magnetic field up to 15\%, while the temperature and line-of-sight velocity are about 20\% worse. This shows how indeed the combination of the two spectral lines provides additional constraints on the physical quantities.
% (100-85)/100*100 = 15%
Treating the macroturbulent velocity as a free parameter increases the errors substantially. The errors increase by approximately 70\% in temperature, 30\% in the magnetic field, and 30\% in the velocity. Although the macroturbulent velocity as a free parameter could reproduce better the observed profiles, the inversion process is more ill-posed because of the degeneracy between the macroturbulent velocity and the physical quantities. One would expect to increase the errors in the temperature and magnetic field strength because all three control the broadening of the spectral lines. However, changing the temperature stratification will also impact the formation region of the spectral line, changing the locations in height where specific velocities are needed.
Finally, the case with the highest impact is the low spectral resolution. The errors in the magnetic field strength increase by 130\%, while the line-of-sight velocity up to 60\%. The error in the temperature is the least affected, with an increase of 12\%. This is expected because the degradation is especially effective in removing the small scales in the polarization profiles because of their oscillatory nature around zero, while Stokes $I$ is barely affected  (see Fig.~\ref{fig:pixel_degradation}).    
% (117-50)/50*100 = 134%
% (0.32-0.2)/0.2*100 = 60%
% (95-85)/85*100 = 11.8%

We can conclude that the inversion process is very sensitive to the following effects (in decreasing order of importance): low-spectral resolution, complete ignorance of the LSF, using only one spectral line, and approximating the LSF with a Gaussian.

\begin{figure}[t!]
\centering
\includegraphics[width=\linewidth,trim={0cm 0cm 0cm 0cm},clip]{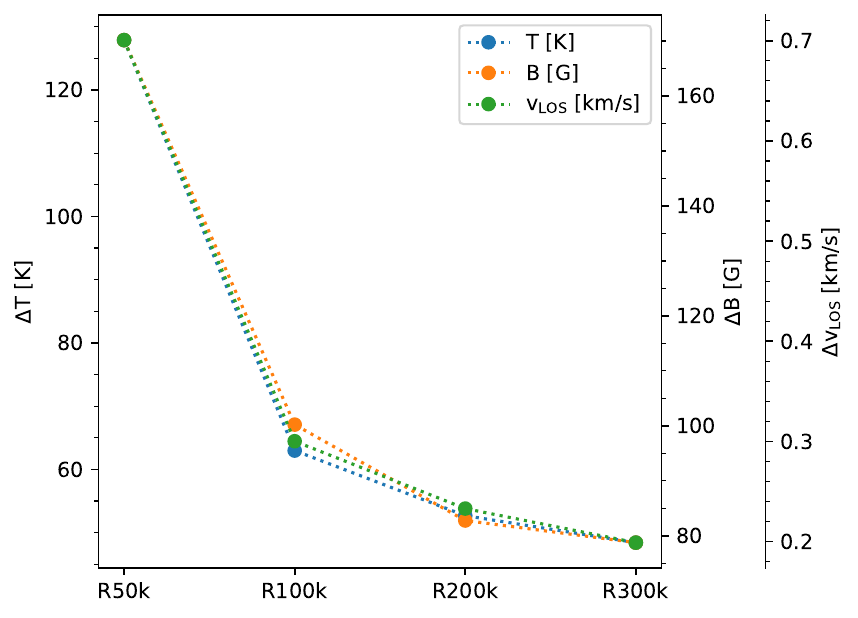}
\caption{Average error in the temperature, magnetic field strength, and line-of-sight velocity for the most realistic case (degraded, noised and resampled) using configuration 2 for different spectral resolutions.}
\label{fig:allresolutions}
% This figure is made using the est_spectralres/4_AnalyzingInversions_summary_noparams.ipynb
\end{figure}

\subsection{Effect of spectral resolution on inversions}

\begin{figure*}[htp!]
\centering
\includegraphics[width=1.0\linewidth]{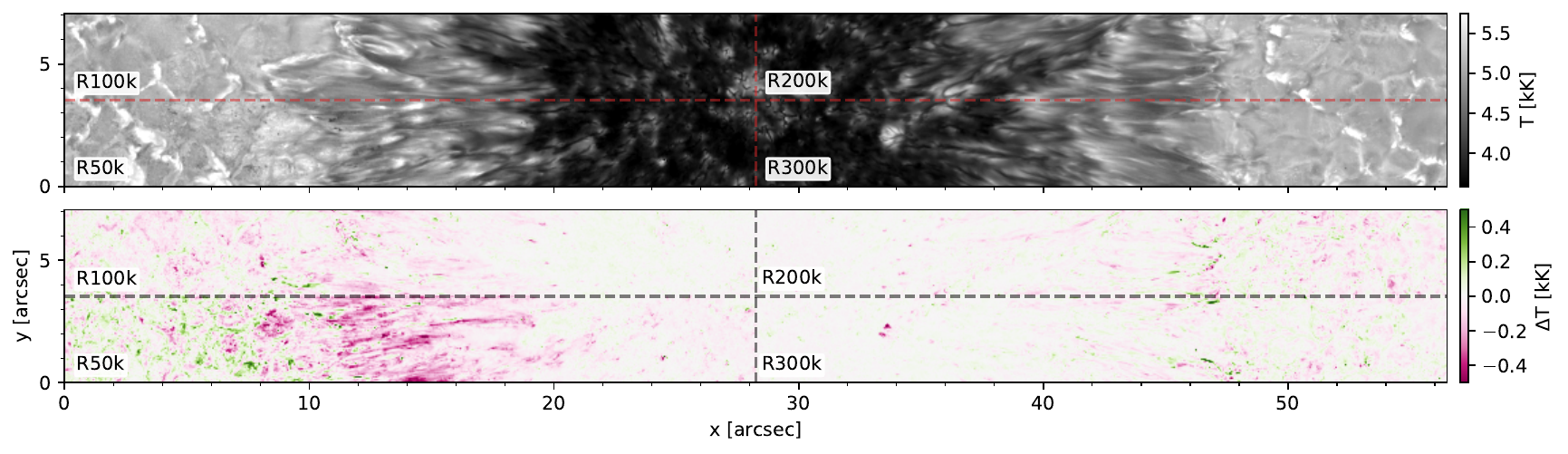}
\includegraphics[width=1.0\linewidth]{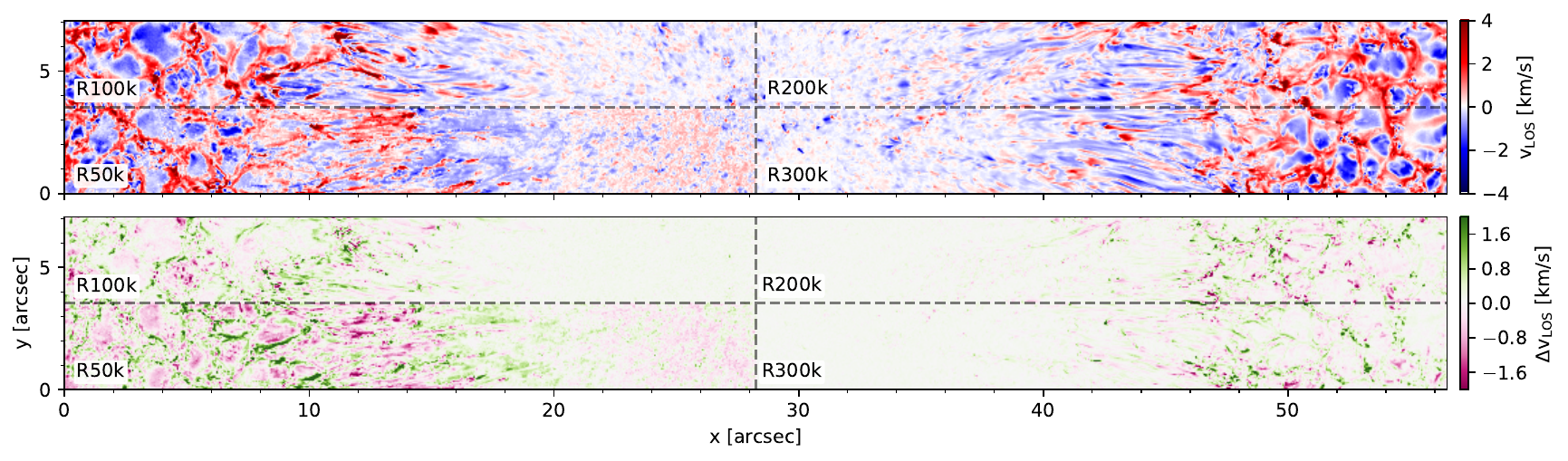}
\includegraphics[width=1.0\linewidth]{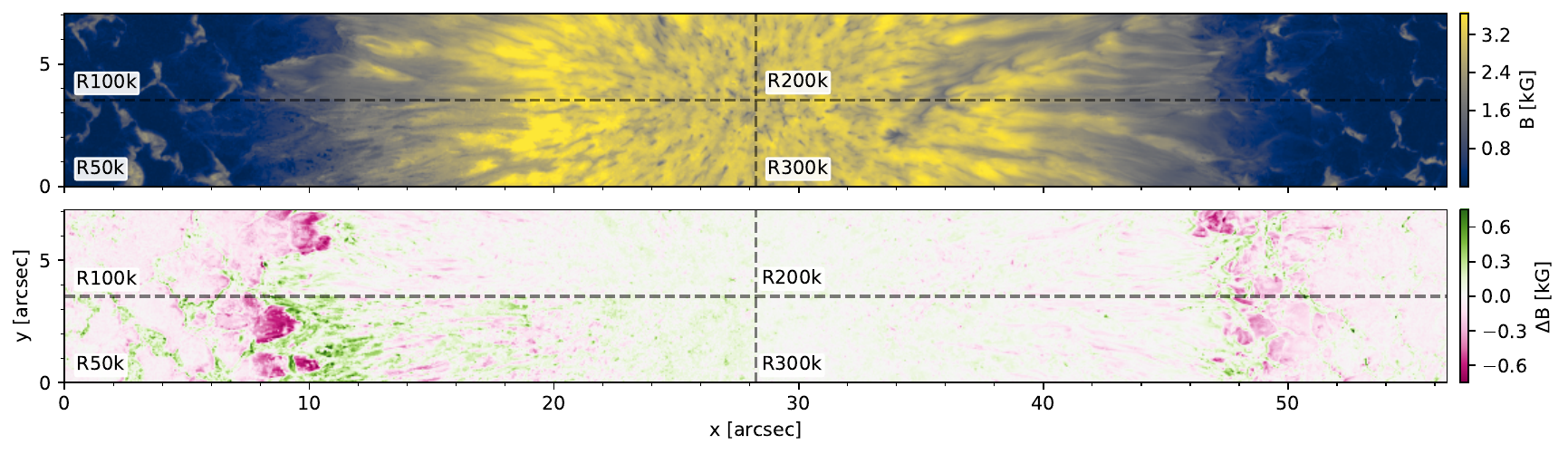}
\caption{Overview of the inferred physical parameters and the difference between the simulation and inversion at $\log \tau_{500}=-1$. The results for different spectral resolutions are shown in every corner of each map. From top to bottom: temperature, line-of-sight velocity, magnetic field strength. Positive difference implies underestimation and negative overestimation of the given physical parameter.} 
\label{fig:invmap}
\end{figure*}

Finally, to isolate the effect of spectral resolution and its impact on the inferred atmospheres, we have repeated the same analysis for four different spectral resolutions: $R=5\times10^4$ to $R=3\times10^5$. The data was degraded using the most realistic scenario, that is: spectrally convolved, noised, and resampled according to the Nyquist-Shannon theorem, and then inverted using the robust configuration (configuration 2). This specific choice of instrumental degradation is intended to make the comparison more realistic and provide information that is better applicable to the design of real-life instruments. The results are shown in Fig.~\ref{fig:allresolutions}. All the errors decrease drastically when the spectral resolution goes from $R=5\times10^4$ to $R=10^5$, despite the increase in noise per wavelength bin. Moreover, a quick comparison of the results at $R=5\times10^4$ between Fig.~\ref{fig:onlyLine_macroturbulence} and Fig.~\ref{fig:allresolutions} shows that the impact of the sampling according to the Nyquist-Shannon theorem is more significant the lower the spectral resolution is. After that, the improvement is less significant and the curve seems to saturate. Note that even ideal inversion with infinite spectral resolution exhibits errors when compared to the simulation, due to the inability of the SIR code to exactly reproduce complicated atmospheric depth stratification found in the simulation. A more detailed comparison is given in Fig.\,\ref{fig:invmap}, where we show the spatial distribution of the inferred parameters and the spatial distribution of the differences between the different inversions and the original simulation. The increase in accuracy between $R=5\times10^4$ and $R=10^5$ is again, evident. The difference between the inferred and original physical parameters depends on the inverted feature: the umbra is typically well reproduced, while the penumbra and quiet Sun inversions exhibit systematic differences. We interpret these systematic differences as a feature of the inversion code, and it is highly possible that different inversion codes or even configurations would result in different systematics. 

From this test, we can conclude that the accuracy improvements brought by spectral resolutions above $10^5$ are minimal. Consequently, expanding the wavelength range to include additional diagnostics may prove to be a more advantageous strategy. The latter approach has the potential to offer better insights into the stratification of physical quantities and could be more cost-effective.

% -------------------------------------------------------
\section{Summary and conclusions}\label{sec:conclusions}
% -------------------------------------------------------

In this study, we analyzed the impact of spectral degradation on the information content in solar spectra. We achieved this by calculating photospheric spectra from a state-of-the-art RMHD simulation of a sunspot and then degrading the Stokes profiles to different spectral resolutions. We have then analyzed the effect of spectral resolution by quantifying the complexity of the Stokes profiles using PCA, the spectral scales across the Stokes profiles using Wavelet decomposition, and the accuracy of the inferred physical parameters using spectropolarimetric inversions. We analyzed a set of specific scenarios because the full parameter space (number and spectral lines of interest, spectral resolution, noise level, LSF functional form, node configuration, etc) is too large to be investigated at once. 

From the study of the dimensionality and spectral scales, we conclude that most of the complex profiles are found in the penumbra and intergranular lanes, which are the regions with intermediate magnetic fields and strong gradients in velocity. They are also the most affected by the degradation. On the other hand, within granules, profiles show features with smaller wavelength scales but their amplitudes are challenging to detect. Finally, profiles from the umbra are less complex in general because, as the convection is inhibited, the stratification tends to be simpler and the broadening due to the magnetic field makes the profiles show spectral scales much larger than the width of the LSF. 

From the analysis of the spectropolarimetric inversions, the model complexity has a strong impact on the inversion results. This is found when the configuration with the highest number of nodes goes from providing the best results in the original case, to the worst as soon as the spectra are degraded and noise is included. Thus the spectral degradation makes the inversion process more ill-posed. This is particularly important for spectropolarimetric inversions, where the complexity of the model should be chosen carefully to avoid overfitting. To address this problem, for example, \cite{AsensioRamos2012ApJ...748...83A} proposed using Bayesian evidence ratios or simple proxies such as the Bayesian Information Criterion \citep[BIC;][]{GideonSchwarz101214aos1176344136} to compare quantitatively different models and favor more complex ones only when they remarkably improve the fit \citep{Sasso2011A&A...526A..42S,DiazBaso2019A&A...625A.129D}. Nowadays, however, the new approach in some codes (STiC, SNAPI, FIRTEZ) consists of allowing a higher number of nodes but limiting the effective degree of freedom with a regularization term in the merit function which will penalize strong gradients (both in the vertical and horizontal direction) if they are not needed when reproducing the observations \citep{2024arXiv240905156D}.

Spectral sampling is also crucial and while in binned data the noise amplitude per wavelength bin is lower, the recovery of the physical parameters is more difficult. According to this experiment, having an over-sampled spectrum might be more beneficial than having a higher signal-to-noise ratio per wavelength bin. This is, of course, in tension with wavelength range, as the coarser wavelength binning allows wider wavelength ranges for the same detector size \citep{Tino2019A&A,Trelles2021ApJ}. Indeed, the combination of the two spectral lines with an approximated LSF provided better results than only one line with a perfect knowledge of the LSF. Given these results, we believe that the LSF can be inferred from the observations, but only if the LSF is coupled (a unique functional form across the FoV) as a pixel-wise version is a very degenerated problem as shown here. This spatially-coupled strategy has shown successful results in determining atomic parameters in spectropolarimetric inversions by \citet{Vukadinovic2024A&A}.

We can conclude that the extent of information loss due to spectral degradation relies on multiple factors (such as spectral resolution, noise level, LSF functional form, physical properties of the solar atmosphere, and degree of freedom in our inversion method, among others). To mitigate this loss, we can incorporate a good estimation of the LSF into the inversion process, avoid coarse samplings (e.g. post-factum spectral binning), and consider including different spectral lines that may compensate for these effects. Similar studies should be conducted under typical chromospheric conditions, where the magnetic field is generally weaker and signals are easily obscured by noise \citep{DiazBaso2019A&A...629A..99D,Yadav2021A&A...649A.106Y}. Consequently, additional studies, akin to this one, should be carried out to optimize the design of new instrumentation according to our scientific requirements \citep{2019arXiv191208650S}. Additionally, it is worth noting that spatial degradation has a more pronounced impact on information loss \citep[e.g.][]{Centeno2023ApJ, Milic2024A&A}. However, the advent of new-generation telescopes with larger apertures and improved adaptive optics systems will yield high-spatial resolution observations where spectral degradation becomes the primary source of information loss. This is likely even more critical when detecting subtle signatures of scattering polarization and Hanle effect \citep{Zeuner2020ApJ, Centeno2022ApJ}. Finding ways to mitigate this loss becomes crucial for the accurate interpretation of such observations.

% -------------------------------------------------------
\begin{acknowledgement}
%{We would like to thank the anonymous referee for their comments and suggestions.}
% General
We thank Juan Manuel Borrero for his comments and suggestions regarding the interpretation of the dimensionality of the spectra. We thank Markus Schmassmann and Matthias Rempel for providing the MURaM simulation of a sunspot, and the Science Advisory Group (SAG) of the European Solar Telescope (EST) for useful discussions and design questions that motivated this project. 
This research is supported by the Research Council of Norway, project number 325491, % ISSRESS (Luc & Carlos)
and through its Centers of Excellence scheme, project number 262622. % RoCS
IM acknowledges the funding provided by the Ministry of Science, Technological Development and Innovation of the Republic of Serbia through the contract 451-03-66/2024-03/200104.
We acknowledge the community effort devoted to the development of the following open-source packages that were used in this work: NumPy (\url{numpy.org}), Matplotlib (\url{matplotlib.org}), SciPy (\url{scipy.org}), and Astropy (\url{astropy.org}).
This research has made use of NASA's Astrophysics Data System Bibliographic Services.
\end{acknowledgement}

\bibliography{references}

\clearpage
\appendix
\section{Dimensionality of the simulation}
\label{app:app1}

\begin{figure*}[htp!]
\centering
\includegraphics[width=0.9\linewidth]{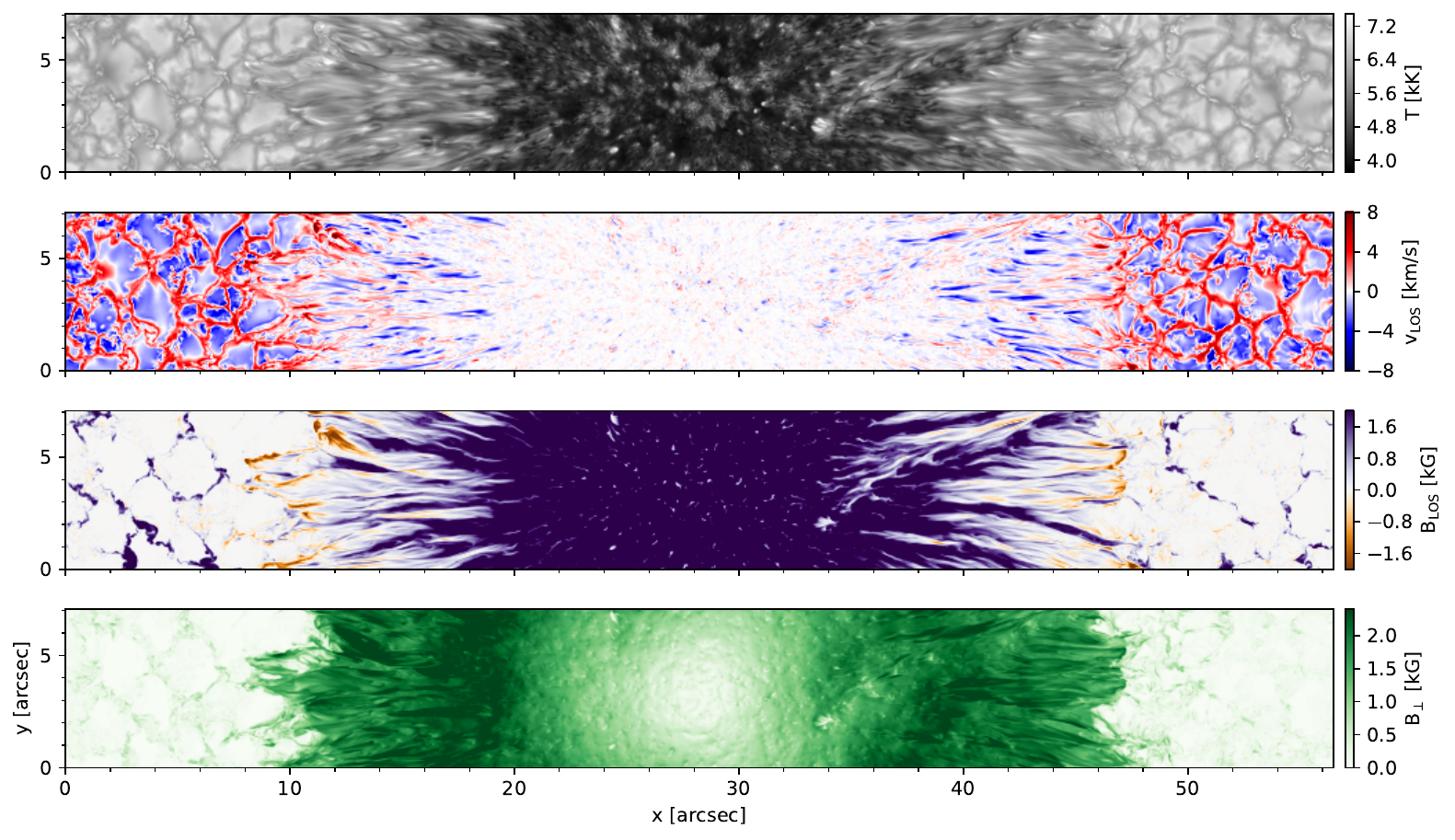}
\caption{Overview of the physical parameters of the simulation at $\tau_{500}=1$. From top to bottom: temperature, line-of-sight velocity, longitudinal magnetic field and transverse magnetic field.} 
\label{fig:simulation}
\end{figure*}

\begin{figure*}[htp!]
\centering
\includegraphics[width=0.9\linewidth]{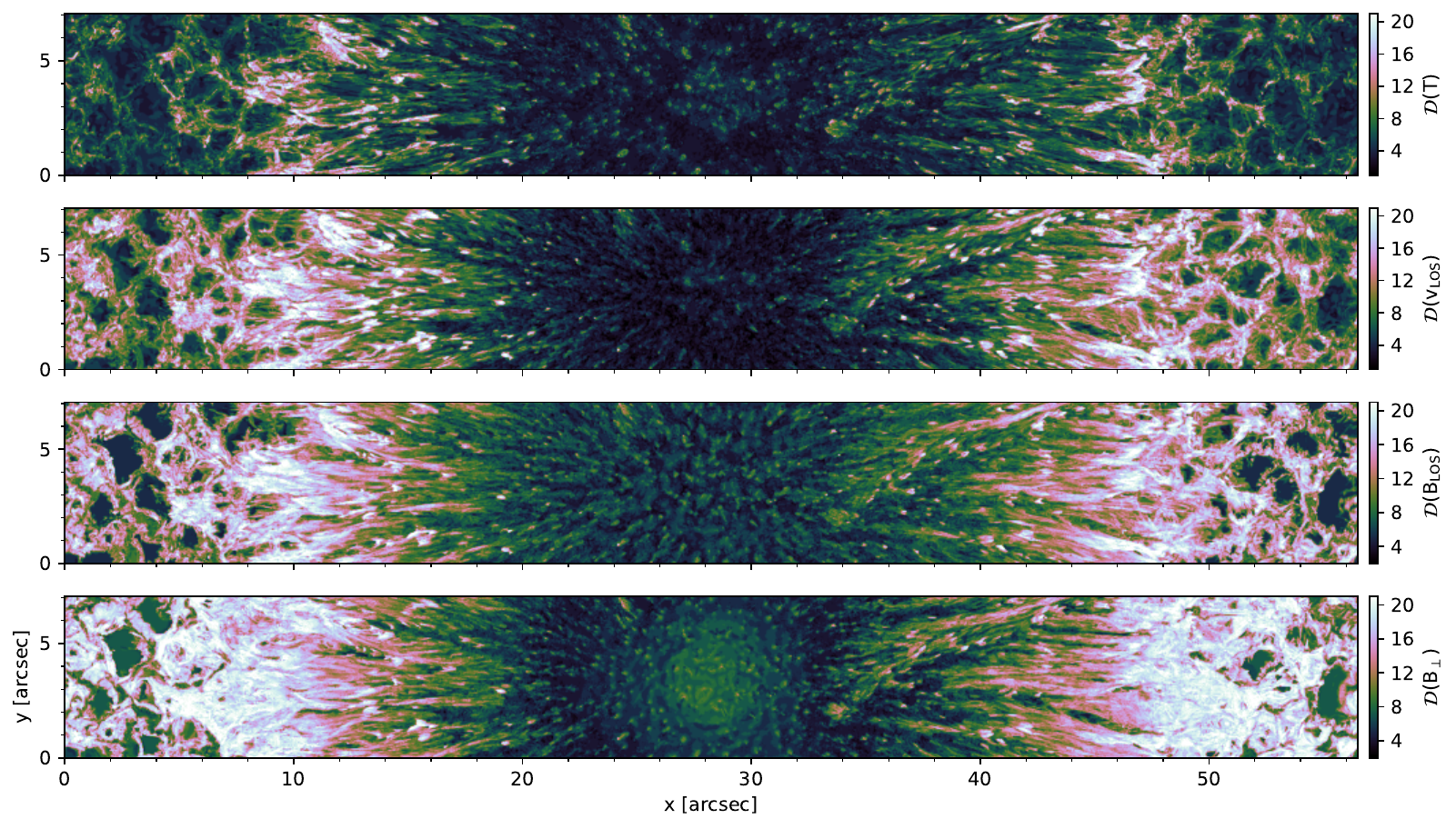}
\caption{Dimensionality of each physical parameter of the simulation in the range $\log\tau_{500}=[0,-2]$ calculated using PCA and the following thresholds: 5~K for the temperature, $10^{-2}$~km/s for the velocity, and 1~G for the magnetic field components.}
\label{fig:dimensionalitysimulation}
\end{figure*}

More complex depth stratifications of temperature, velocity, and the magnetic field vector, will result in more complex shapes of the Stokes profiles. This means that the dimensionality of the Stokes profiles is related to the dimensionality of the underlying atmosphere. To complement our results from Section\,\ref{sec:pca}, we show the dimensionality of the depth stratifications of the temperature, velocity, and magnetic field in our model atmosphere. We calculate the dimensionality following the same PCA approach as in Section\,\ref{sec:pca} (see Eq.\,\ref{eq:pca}), where now the basis vectors are functions of optical depth. The dimensionality is calculated over the range of optical depths where the \ion{Fe}{i} lines are sensitive to the stratification, i.e., $\log\tau_{500}=[0,-2]$. The threshold criteria used to determine this dimensionality are 5~K for the temperature, 0.01~km/s for the velocity, and 1~G for the magnetic field components, where we treat horizontal and vertical magnetic fields separately. Fig\,\ref{fig:simulation} shows the spatial distribution of these four physical parameters at the optical depth $\tau_{500}=1$, while Fig.\,\ref{fig:dimensionalitysimulation} shows the spatial distribution of the dimensionality of the quantities mentioned above.

%-----------------------------------------------------------

\section{MPySIR: a parallel wrapper for SIR}
\label{app:app2}

As the original SIR code\footnote{Available at \url{https://github.com/BasilioRuiz/SIR-code}} \citep{SIR} is not parallelized and the inversion process of large datasets is computationally expensive, we have implemented a parallelized version of the SIR code, called MPySIR\footnote{Available at \url{https://github.com/cdiazbas/MPySIR}.}. The code is written in Python and uses the MPI library to distribute the inversion task across multiple processors. This implementation does not modify the original SIR implementation, and all the MPI calls are done from Python. 

This new implementation integrates in a single configuration file the previous functionalities of the SIR code and new functionalities related to the parallelization. Through this file, users can control various aspects, such as the input/output files, abundances, the mode of synthesis or inversion, the number of nodes for each physical parameter, and more. Additional features include debugging tools, the option to perform inversions only within a specified region of the dataset, the ability to combine different inversion results, the option to use previous inversion results as inputs for subsequent cycles, along with numerous other possibilities. The only feature that we did not carry over is multi-component inversions, mostly because we are interested in very high-resolution observations where we deem that feature unnecessary. 

\end{document}